\documentclass[aps,prl,preprint,tightenlines,superscriptaddress,byrevtex]{revtex4}
\usepackage{graphicx} 
\usepackage{dcolumn}  
\usepackage{enumerate}
\usepackage{csquotes}
\graphicspath{{ps}}

\begin{document}



\title{ \quad\\[0.5cm]  Latest results on $B\to DK/D\pi$ decays from Belle}


\author{P. K. Resmi\\ Indian Institute of Technology Madras, Chennai, India \\ (On behalf of the Belle Collaboration)}



\begin{abstract}
The CKM angle $\phi_3$ is less precisely known than the angle $\phi_1$ and the only one that is accessible with tree-level decays in a theoretically clean way. The key method to measure $\phi_3$ is through the interference between $B^+\to D^0 K^+$ and $B^+ \to \bar D^0 K^+$ decays which occurs if the final state of the charm-meson decay is accessible to both the $D^0$ and $\bar D^0$ mesons. To achieve the best sensitivity, a large variety of $D$ and $B$ decay modes is required, which is possible at Belle experiment as almost any final state can be reconstructed including those with photons. The results from Belle, as well as the ongoing studies, are discussed here. The details of the planned measurement at Belle II, a substantial upgrade of the Belle detector which will operate at the SuperKEKB energy-asymmetric $e^+ e^-$ collider, are also discussed. The results from the first collisions at Belle II during April-July 2018 are shown too.
\end{abstract}


\maketitle

\tighten


\section{Introduction}

\begin{figure}[ht!]
\centering
\includegraphics[width = 0.7\columnwidth]{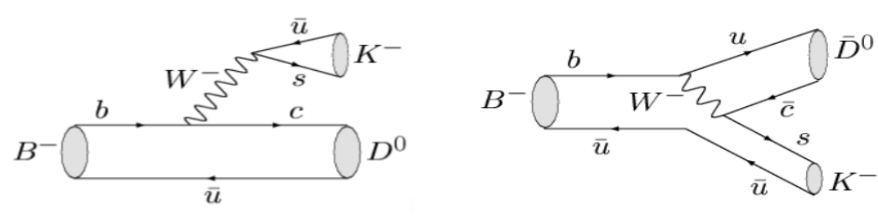}  \caption{Color-favored (left) and -suppressed (right) $B^{-} \to DK^{-}$ processes.} \label{fig:diagram}
\end{figure}

Among the three CKM \cite{C,KM} angles, the uncertainty on $\phi_3$ is much worse than that on $\phi_1$ \cite{PDG}. This is because of the small branching fractions of decays sensitive to $\phi_3$. An improved measurement of $\phi_3$ is essential for testing the standard model description of $\mathit{CP}$  violation. The color-favored $B^{-} \to D^{0} K^{-}$ and color-suppressed $B^{-} \to \bar{D^{0}}K^{-}$ decays, where $D$ indicates a neutral charm meson reconstructed in a final state common to both $D^{0}$ and $\bar{D^{0}}$, provide $\mathit{CP}$-violating observables, that are sensitive to $\phi_3$. The Feynman diagrams are shown in Fig.~\ref{fig:diagram}. These are tree-level decays and hence the theoretical uncertainty is $\mathcal{O}(10^{-7})$ \cite{Brod}. 

If the amplitude for the color-favored decay is  $A_{\rm fav} = A$, then the color-suppressed one can be written as $A_{\rm sup} = Ar_{B}e^{i(\delta_{B} - \phi_{3})}$, where $\delta_{B}$ is the strong phase difference between the decay processes, and 
\begin{equation}
r_{B} = \frac{\mid A_{\rm{sup}} \mid} {\mid A_{\rm{fav}}\mid}.
\end{equation}
The statistical uncertainty on $\phi_3$ is proportional to $r_B$. For $B \to DK$ decays, $r_{B} \sim 0.1$, whereas for $B\to D\pi$, it is $0.05$. Though $B \to D \pi$ decays are not very sensitive to $r_{B}$ and $\phi_{3}$, they serve as excellent control sample modes for $B \to DK$ to estimate systematic uncertainties due to their similar kinematics.

\section{Methods for $\phi_{3}$ extraction}
 \label{sec:methods}

There are different methods to extract $\phi_3$ that are classified according to the $D$ meson final state:
\begin{enumerate}[(i)]
\item {\bf GLW} \cite{GLW} method: $CP$ eigenstates such as $K^{+}K^{-}$, $\pi^{+}\pi^{-}$, $K_{\rm S}^{0}\pi^{0}$,
\item {\bf ADS} \cite{ADS} method: doubly-Cabibbo-suppressed states such as $K^{+}X^{-}$, where $X^{-}$ can be $\pi^{-}, \pi^{-}\pi^{0}, \pi^{-}\pi^{-}\pi^{+}$, and
\item {\bf GGSZ} \cite{GGSZ} method: self-conjugate multibody states such as $K_{\rm S}^{0}\pi^{+}\pi^{-}$, $K_{\rm S}^{+}K^{+}K^{-}$, $K_{\rm S}^{0}\pi^{+}\pi^{-}\pi^{0}$.
\end{enumerate}

$\phi_{3}$-sensitive parameters can be extracted by taking a ratio between the suppressed and favored decay rates and a measurement of $CP$ asymmetries between them in both GLW and ADS methods. Four GLW parameters $R_{CP}^{\pm}~=~1 + r_{B}^{2} \pm 2 r_{B} \delta_{B} \cos \phi_{3}$ and $A_{CP}^{\pm}~=~\pm 2 r_{B} \sin \delta_{B} \sin \phi_{3}/R_{CP}^{\pm}$, where the $+$ and $-$ indicate a $CP$-even and $CP$-odd eigenstate, respectively and two ADS parameters $R_{\rm{ADS}}~=~r_{B}^{2} + r_{D}^{2} + 2 r_{B} r_{D} \cos(\delta_{B} + \delta_{D})\cos\phi_{3}$ and $A_{\rm{ADS}}~=~2 r_{B}r_{D} \sin(\delta_{B} \delta_{D})\sin \phi_{3}/R_{\rm{ADS}}$ are obtained for extracting $\phi_{3}$. Here, $r_{D}$ and $\delta_{D}$ are the ratio of the amplitudes of the suppressed and favored $D$ decays, and the $D$ strong phase, respectively. These are external inputs from charm measurements. For multibody $D$ decays, additional inputs like $CP$-content $F_{+}$ and coherence factor $\kappa$ are needed for GLW and ADS methods, respectively.

In GGSZ method, the Dalitz space is binned into regions with differing strong phases, which allows $\phi_{3}$ to be determined from a single channel in a model-independent manner. This eliminates the model-dependent systematic uncertainty in the measurement. The signal yield in each bin is 
\begin{equation}
\Gamma_{i}^{\pm} \propto K_{i} + r_{B}^{2}\bar{K_{i}} +2\sqrt{K_{i}\bar{K_{i}}}(c_{i}x_{\pm} + s_{i}y_{\pm}),
\end{equation} \label{Eg:GGSZ}

\noindent where $x_{\pm} = r_{B}\cos (\delta_{B} \pm \phi_{3})$; $y_{\pm} = r_{B}\sin (\delta_{B} \pm \phi_{3})$. Here, $K_{i}$ and $\bar{K_{i}}$ are the fraction of flavour-tagged $D^{0}$ and $\bar{D^{0}}$ events in the $i^{\rm th}$ bin, respectively, which can be estimated from $D^{*\pm} \to D \pi^{\pm}$ decays with good precision due to their large sample size. The parameters $c_{i}$ and $s_{i}$ are the amplitude-weighted average of the $\cos$ and $\sin$ of the strong phase difference between $D^{0}$ and $\bar{D^{0}}$ over the $i^{\rm th}$ bin; these parameters need to be determined at a charm factory experiment like CLEO-c or BESIII, where the quantum-entangled $D^{0}\bar{D^{0}}$ pairs are produced via $e^{+}e^{-} \to \psi(3770) \to D^{0}\bar{D^{0}}$~\cite{PKR-fpcp}.

\section{Latest results from Belle}
\begin{figure}[ht!]
\begin{tabular}{cc}
\includegraphics[scale=0.4]{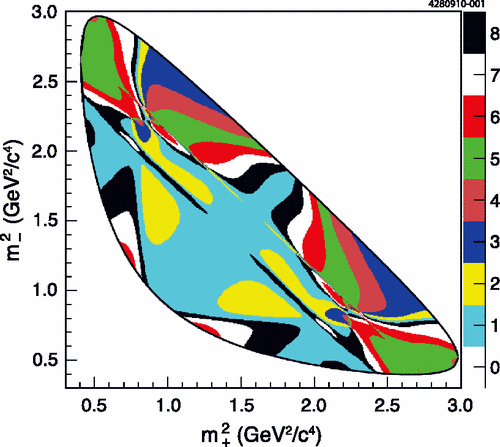}&
\includegraphics[scale=0.48]{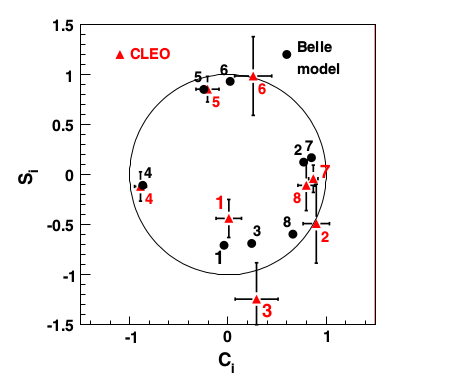} 
\end{tabular}
\caption{Binned Dalitz plot of $D\rightarrow K_{\rm S}^{0}\pi^{+}\pi^{-}$ (left) and $c_i$, $s_i$ results in each bin (right) for $B\to D(K_{\rm S}^0\pi^+\pi^-)K$~\cite{Kspipi}.} \label{fig:KsPiPi}
\end{figure}

The golden mode to determine $\phi_3$ at the $B$-factories, $B\to D(K_{\rm S}^0\pi^+\pi^-)K$, has been analysed via the GGSZ formalism in both model independent~\cite{BELLEGGSZ} and model dependent~\cite{BELLEGGSZ2} ways by Belle. In the model dependent approach, the $D$ Dalitz space is fitted with an amplitude model, which leads to an associated uncertainty of approximately 9$^{\circ}$. In the model independent approach, the model uncertainty is replaced by the statistical uncertainty of the measured $c_i$  and $s_i$ input values. Here the Dalitz plane is binned into different regions guided by the amplitude model for maximum sensitivity as shown in Fig.~\ref{fig:KsPiPi}.

\begin{figure}[ht!]
\centering
\includegraphics[width = 0.5\columnwidth]{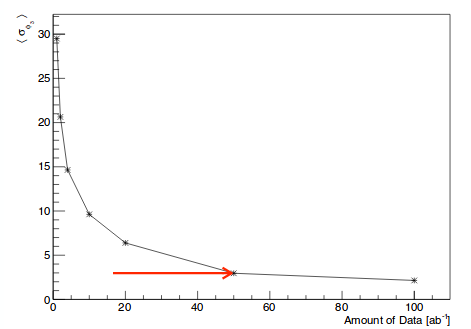}  \caption{$\phi_{3}$ sensitivity using $B\to D(K_{\rm S}^0\pi^+\pi^-)K$ with the amount of data collected at Belle~II. The red arrow indicates the expected sensitivity from the 50~ab$^{-1}$ sample.} \label{Fig:belle2}
\end{figure}

This is the first model-independent Dalitz analysis performed and the result obtained is $\phi_{3}~=~(77^{+15.1}_{-14.9}\pm 4.1 \pm 4.3)^{\circ}$~\cite{BELLEGGSZ}, where the uncertainties are statistical, systematic and from external CLEO-c inputs, respectively. The measurement is limited by statistics. For future model-independent measurements, the inputs from BES III are imperative. A conservative estimate of the expected precision at Belle II with the full data sample of 50~ab$^{-1}$, gives $\sigma_{\phi_3} = 3^{\circ}$ as shown in Fig.~\ref{Fig:belle2}~\cite{PK-belle2}.

\begin{figure}[htb!]
\centering
\includegraphics[width = 0.7\columnwidth]{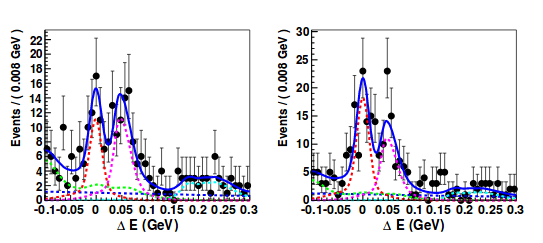}  \caption{$\Delta E$ distributions for $B^{-}$ (left) and $B^{+}$ (right) for $D_{CP+}$ modes.} \label{Fig:D*0K1}
\end{figure}

\begin{figure}[htb!]
\centering
\includegraphics[width = 0.7\columnwidth]{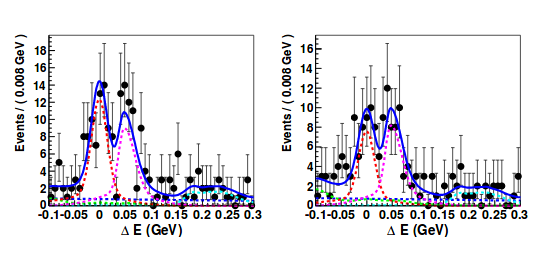}  \caption{$\Delta E$ distributions for $B^{-}$ (left) and $B^{+}$ (right) for $D_{CP-}$ modes.} \label{Fig:D*0K2}
\end{figure}

The GLW ($KK, \pi\pi, K_{\rm S}^0\pi^0, K_{\rm S}^0\eta$) and ADS ($K\pi$) modes with $B\to D^{*0}K$ have been analysed with the full Belle dataset. The energy difference is defined as $\Delta E = E_{B}^{*} - E_{\rm{beam}}^{*}$ where $E_{B}^{*}$ is the energy of the  $B$ meson candidate and $E_{\rm{beam}}^{*}$ is the beam energy, both calculated in the center-of-mass frame. The $\Delta E$ distributions for the GLW modes are given in Fig~\ref{Fig:D*0K1} and \ref{Fig:D*0K2}.
With the $CP$ modes for $D^*\to D\pi^0, D\gamma$ decays combined, the results from the GLW analysis is obtained to be $A_{CP+}$ = $-0.14\pm 0.10(\rm stat)\pm 0.01(\rm syst)$ and $A_{CP-}$ = $+0.22\pm 0.11(\rm stat)\pm 0.01(\rm syst)$. In the ADS method, the results are obtained separately for $D^*\to D\pi^0$ and $D\gamma$ as $R_{D^*K, D\pi^0}$~=~[1.0$^{+0.8}_{-0.7}(\rm stat)^{+0.1}_{-0.2}(\rm syst)$]$\times 10^{-2}$ and $R_{D^*K, D\gamma}$~=~[3.6$^{+1.4}_{-1.2}(\rm stat)\pm 0.2(\rm syst)$]$\times 10^{-2}$, respectively.

The suppressed decay $B\to D(K^{+}\pi^-\pi^0)K$ has been observed with 3.2$\sigma$ significance with the full Belle data~\cite{kpipi0}. The coherence factor $\kappa$ for $D\to K^{+}\pi^-\pi^0$ is close to 1 and hence the dilution due to the strong phase from $D$ decay multi-particle phase space is quite small. $B^0\to D^0K^{*0}$ modes have also been analysed with $D\to K\pi$ (ADS)~\cite{DK*1} and $D\to K_{\rm S}^0\pi^+\pi^-$ (GGSZ)~\cite{DK*2}.

\begin{figure}[ht!]
\begin{tabular}{cc}
\includegraphics[scale=0.3]{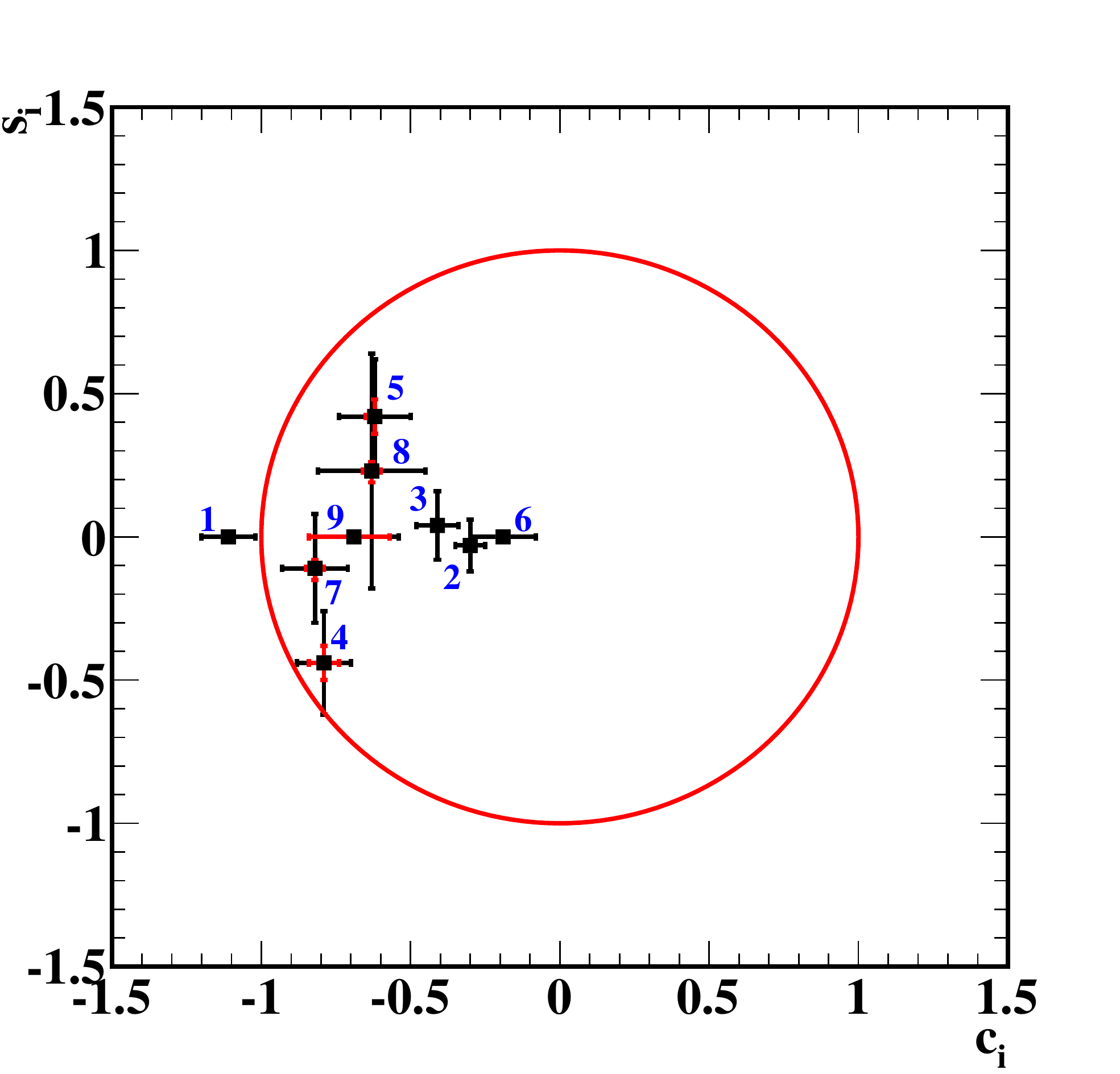}&
\includegraphics[scale=0.42]{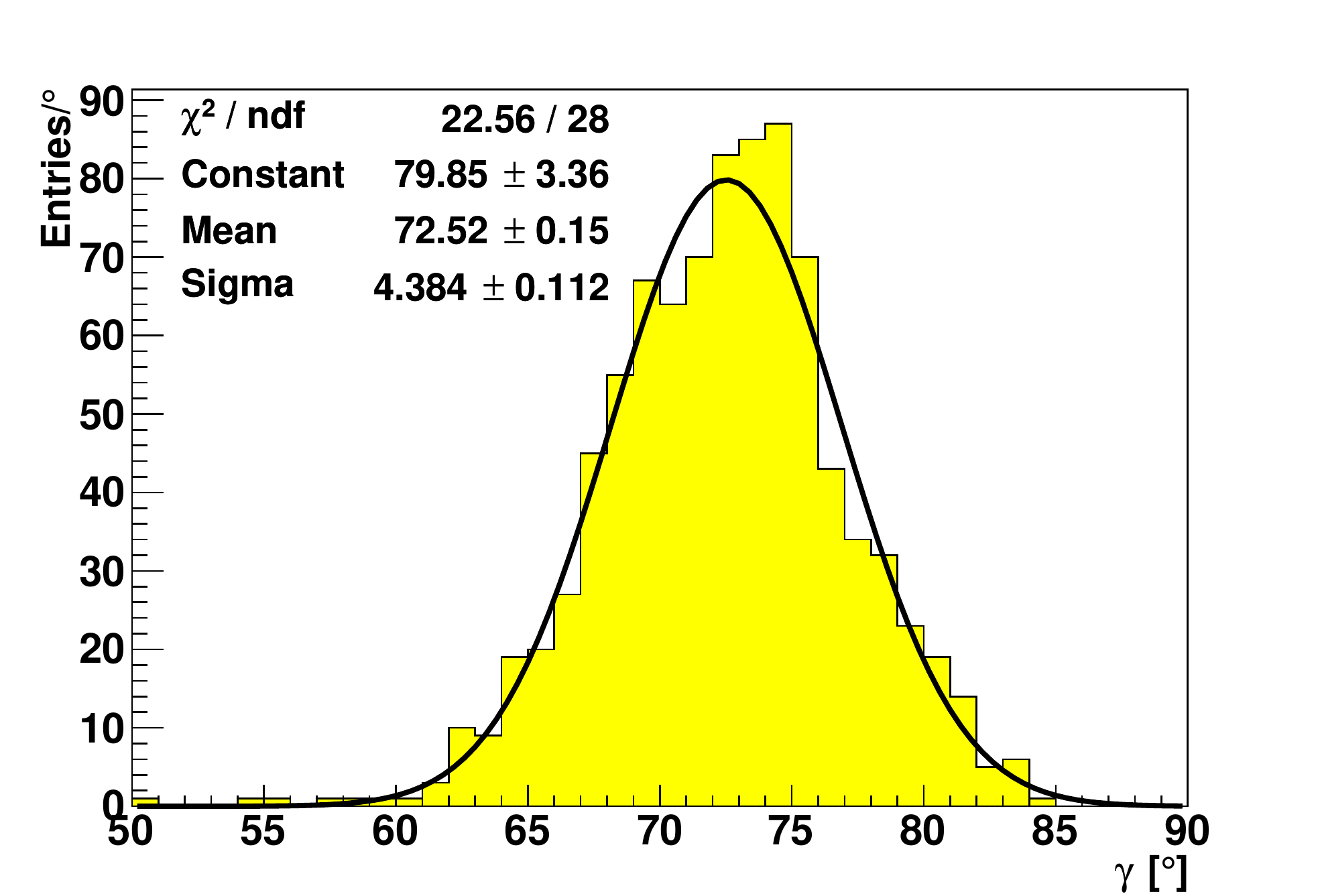} 
\end{tabular}
\caption{$c_i$, $s_i$ results in each bin (right) and the sensitivity with 50~ab$^{-1}$ data sample from Belle II (right) for $B\to D(K_{\rm S}^0\pi^+\pi^-\pi^0)K$.} \label{fig:KsPiPiPi0}
\end{figure}

There are new ongoing analyses with the full Belle dataset. $D\to K_{\rm S}^0\pi^+\pi^-\pi^0$ is a good potential candidate with a relatively large branching fraction of 5.2\%~\cite{PDG}. The decay proceeds through interesting resonance substructures like $K_{\rm S}^{0}\omega$  ($CP$ eigenstate and GLW like), $K^{*-}\pi^{+}\pi^{0}$  (Cabibbo-favored state and ADS like). The strong phase difference parameters $c_i$ and $s_i$ have been measured with CLEO-c data~\cite{PKR}. The $D$ phase space is binned around the different resonances present, in the absence of an amplitude model. We expect $\sigma_{\phi_3} = 25^{\circ}$ with the full Belle dataset and $4.4^{\circ}$ with 50~ab$^{-1}$ data sample from Belle II (Fig.~\ref{fig:KsPiPiPi0}).

Time dependent measurements using $B\to D^*\rho$ are currently under way with the full Belle dataset. There are approximately 60,000 signal events reconstructed. The value of $2\phi_1 + \phi_3$ can be extracted with angular time dependent fit. We foresee $\sigma(2\phi_1 + \phi_3) = 11^{\circ}$ for Belle II at 50~ab$^{-1}$.

\section{Early Belle II results}
\begin{figure}[ht!]
\includegraphics[scale=0.35]{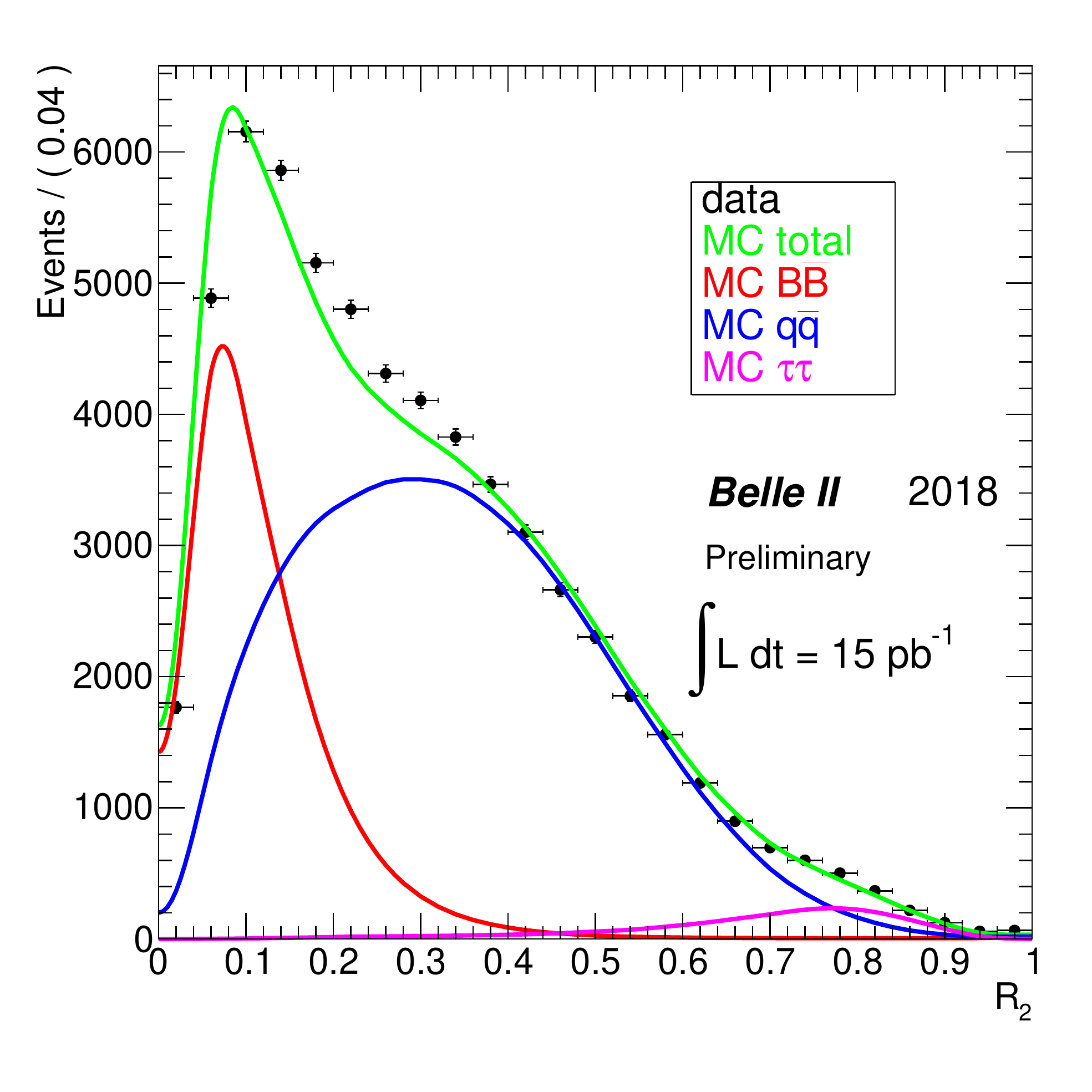}
\caption{$R_2$ distribution from early Belle II data.} \label{Fig:R2}
\end{figure}

The Belle II detector witnessed its first $e^+e^-$ collision on 26 April 2018, which was followed by a run that completed on 17 July 2018. The full vertex detector system was absent during this pilot run. However, one module in each layer was used for background calibrations. About 0.5~fb$^{-1}$ data have been accumulated during this time. It is confirmed that the collisions happen at $\Upsilon (4S)$ by looking at event shape parameters like $R_2$, which is the ratio of second and zeroth Fox-Wolfram moment~\cite{FW}. The distribution is shown in Fig.~\ref{Fig:R2}. For $B\bar{B}$ events, the value is close to zero and for $q\bar{q}$ events, it goes to higher values. The data agrees with the Monte Carlo (MC) expectations and the presence of $B\bar{B}$ suggests the production of $\Upsilon (4S)$.

\begin{figure}[ht!]
\begin{tabular}{cc}
\includegraphics[width=0.5\columnwidth]{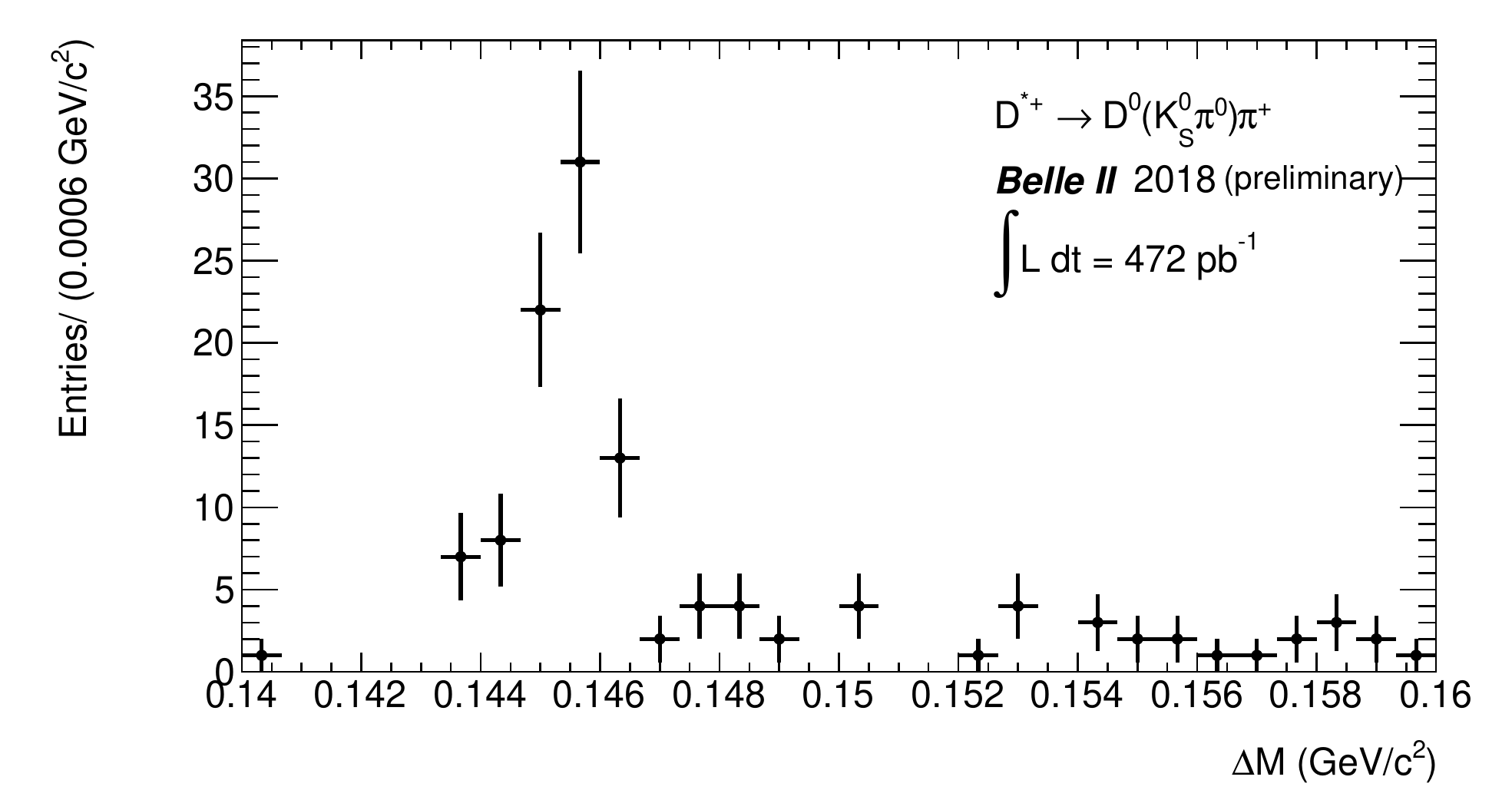} &
\includegraphics[width=0.5\columnwidth]{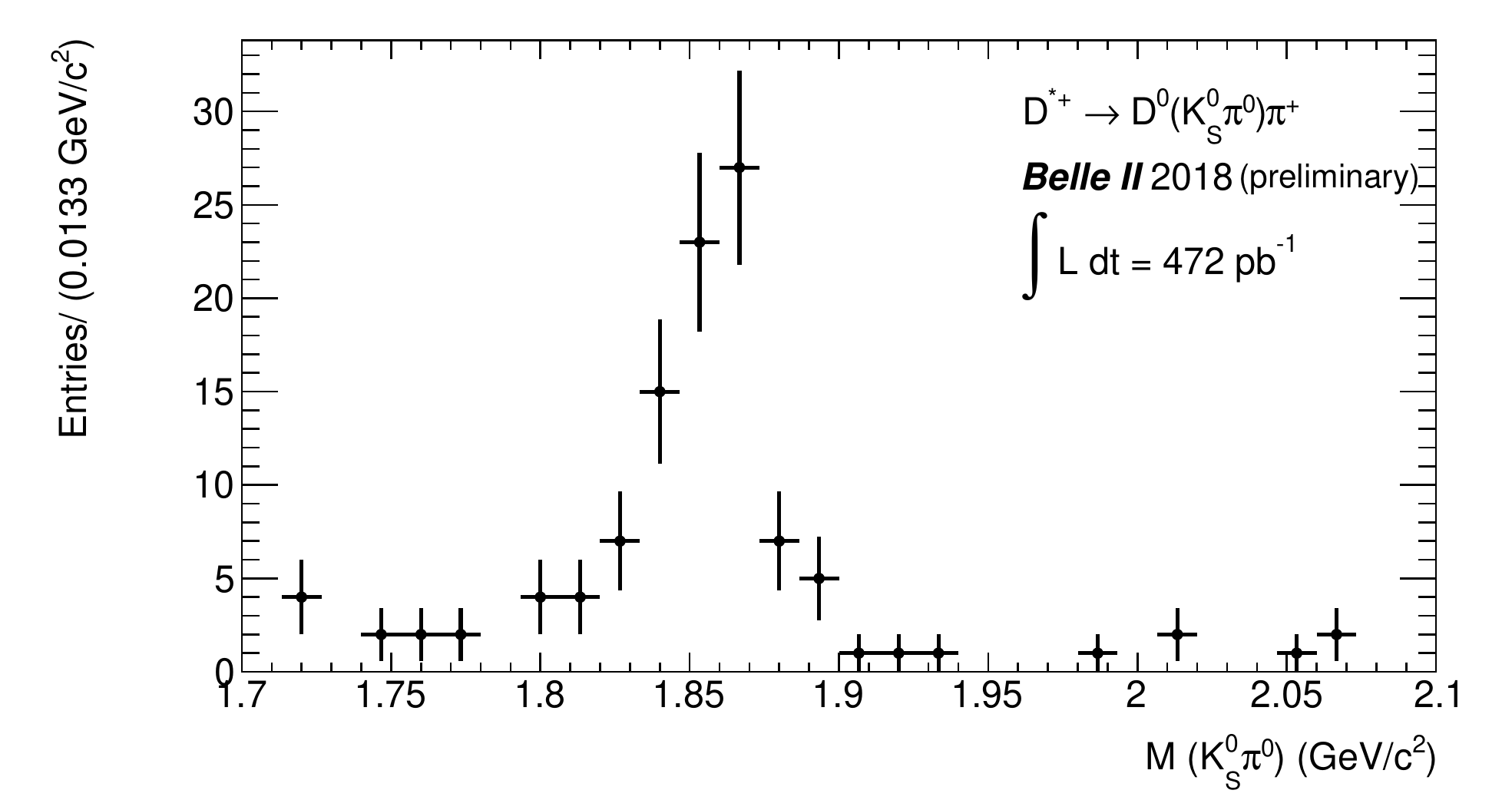} \\
\end{tabular}
\caption{$\Delta M$ (left) and $M_{D}$ (right) distributions for $D^{*\pm} \to D(K_{\rm S}^{0}\pi^{0})\pi^{\pm}_{\rm{slow}}$ decays.}\label{Fig:kspi0}
\end{figure}

\begin{figure}[ht!]
\begin{tabular}{cc}
\includegraphics[width=0.5\columnwidth]{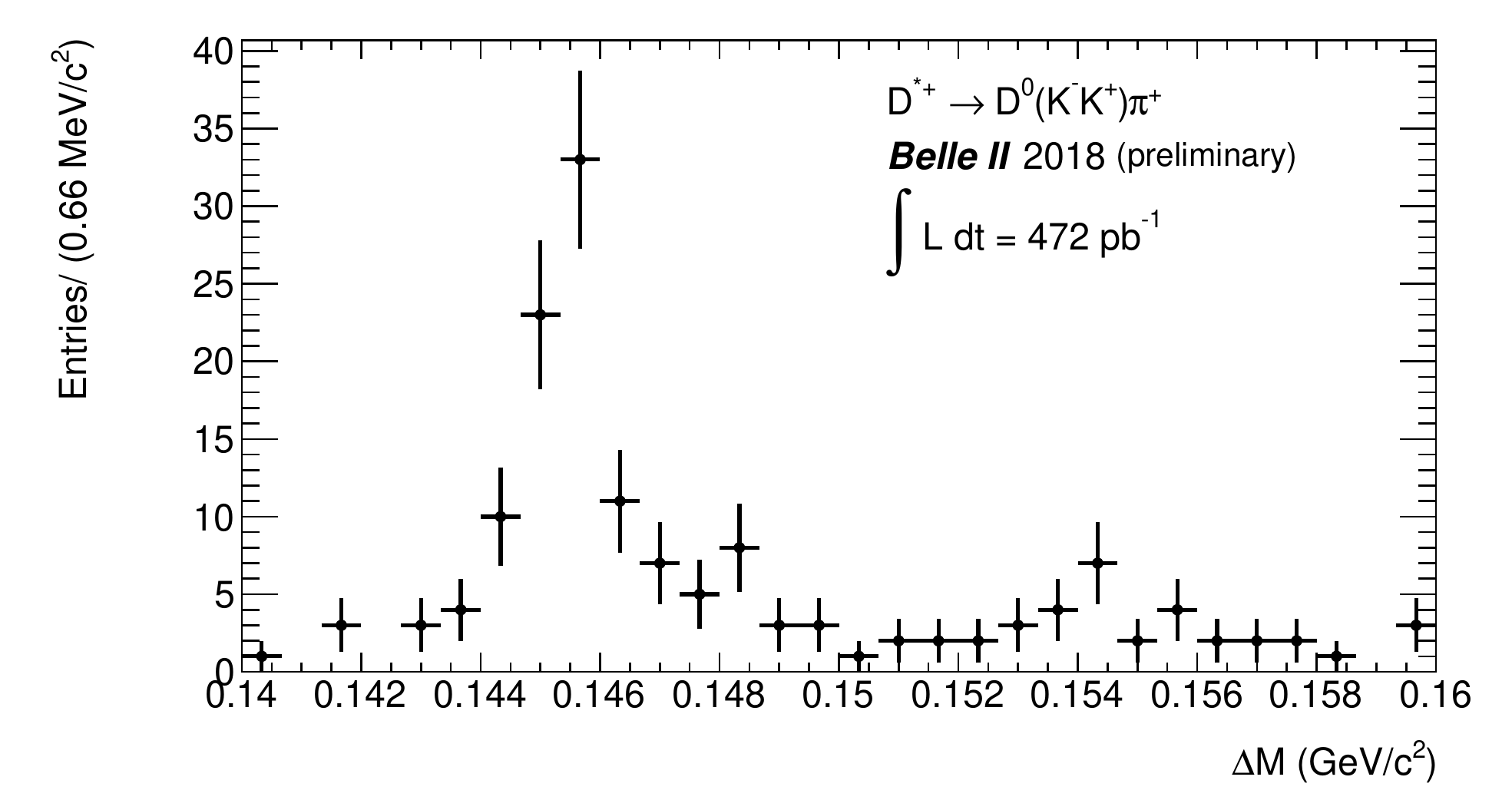} &
\includegraphics[width=0.5\columnwidth]{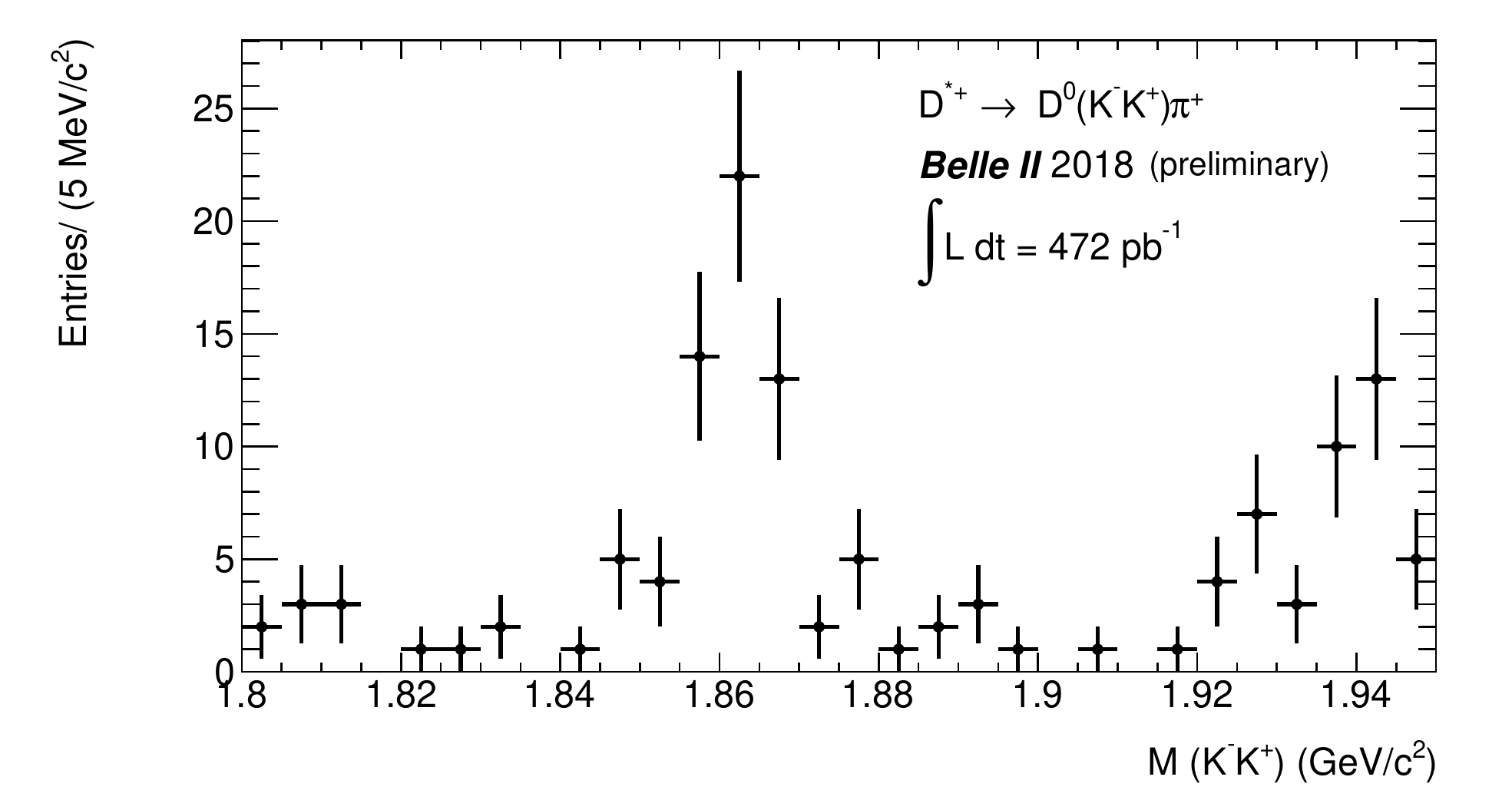} \\
\end{tabular}
\caption{$\Delta M$ (left) and $M_{D}$ (right) distributions for $D^{*\pm} \to D(KK)\pi^{\pm}_{\rm{slow}}$ decays.}\label{Fig:kk}
\end{figure}

Different $D$ decay modes have been rediscovered, including $K_{\rm S}^0\pi^0$, which is a $CP$-odd eigenstate and $KK$, which is $CP$-even as well as being singly-Cabibbo-suppressed. The $\Delta M$, the reconstructed mass difference between $M_{D^*}$ and $M_{D}$, and $M_{D}$ distributions for these modes are given in Fig.~\ref{Fig:kspi0} and \ref{Fig:kk}. The resolution is comparable to Belle II MC expectations. 

\begin{figure}[ht!]
\begin{tabular}{cc}
\includegraphics[width=0.5\columnwidth]{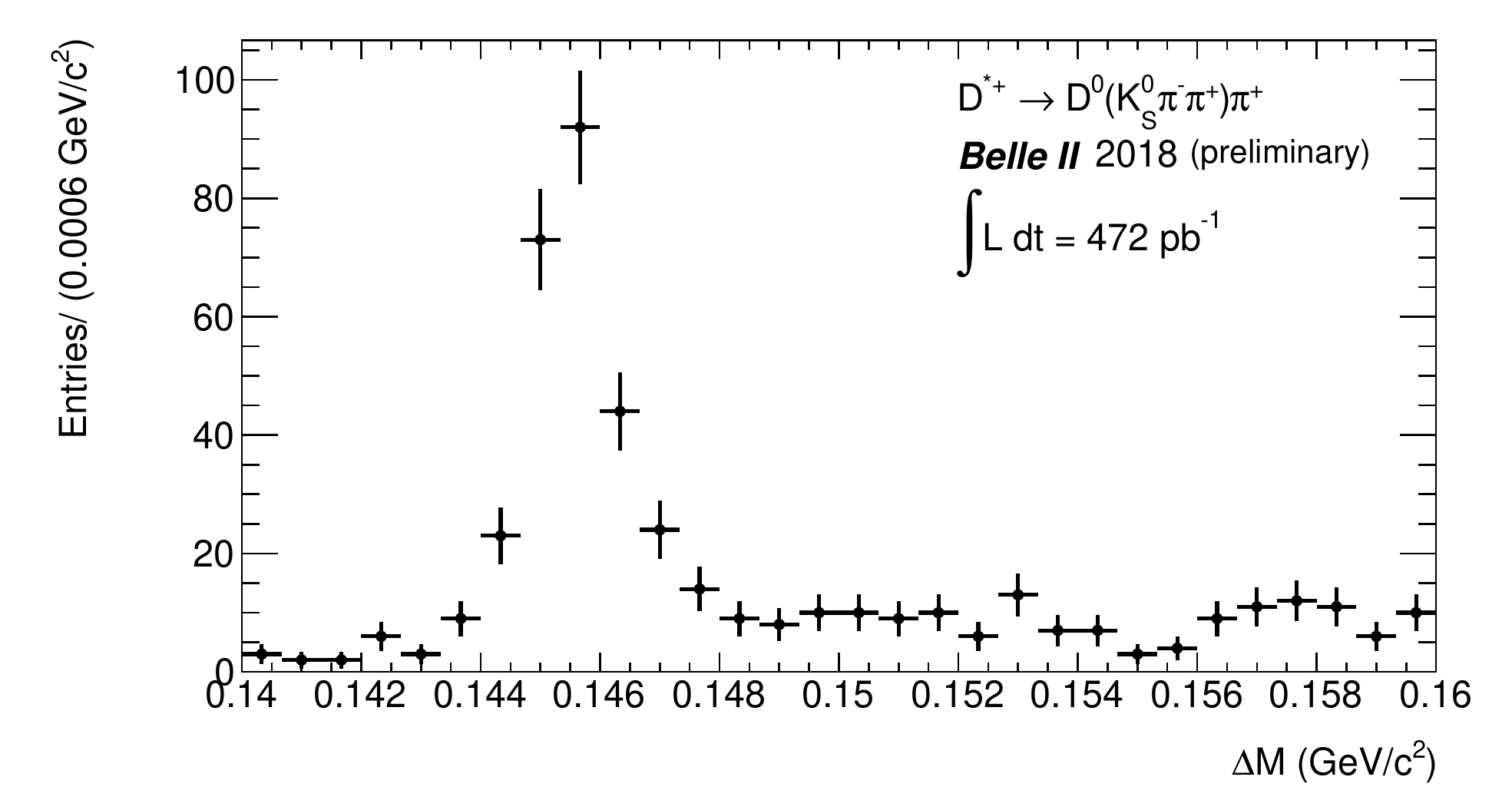} &
\includegraphics[width=0.5\columnwidth]{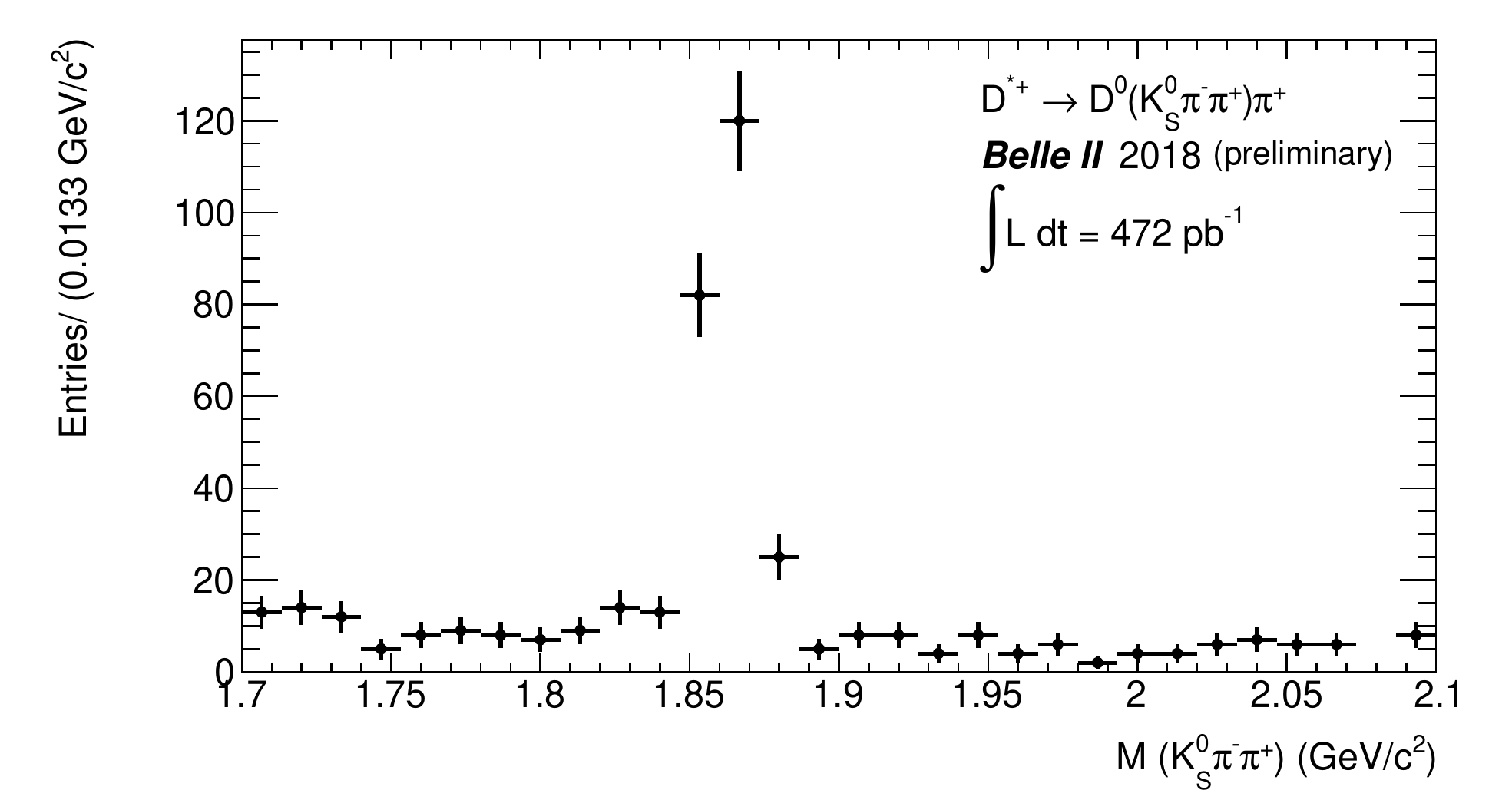} \\
\end{tabular}
\caption{$\Delta M$ (left) and $M_{D}$ (right) distributions for $D^{*\pm} \to D(K_{\rm S}^{0}\pi^{+}\pi^-)\pi^{\pm}_{\rm{slow}}$ decays.}\label{Fig:kspipi}
\end{figure}

\begin{figure}[ht!]
\begin{tabular}{cc}
\includegraphics[width=0.5\columnwidth]{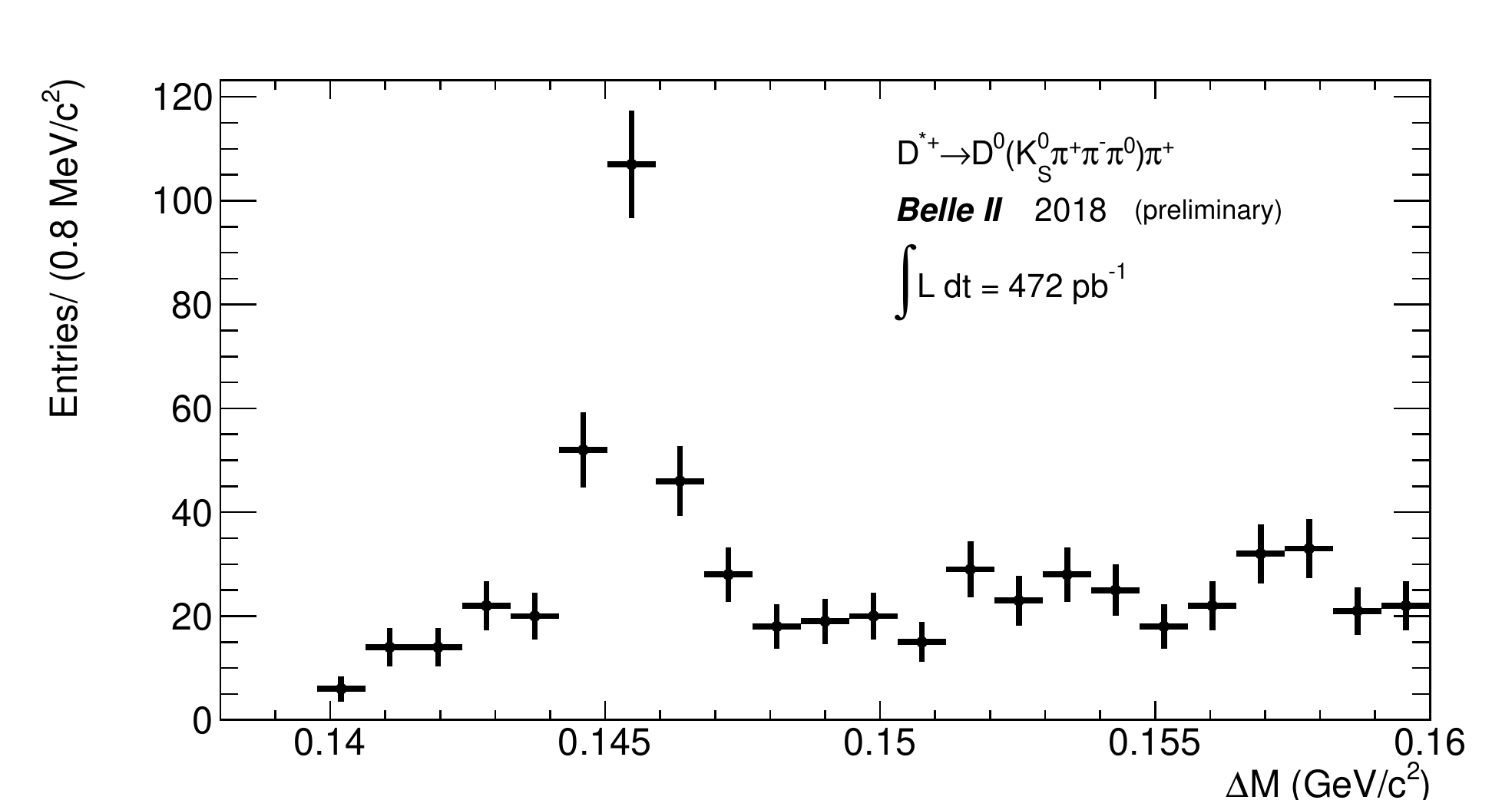} &
\includegraphics[width=0.5\columnwidth]{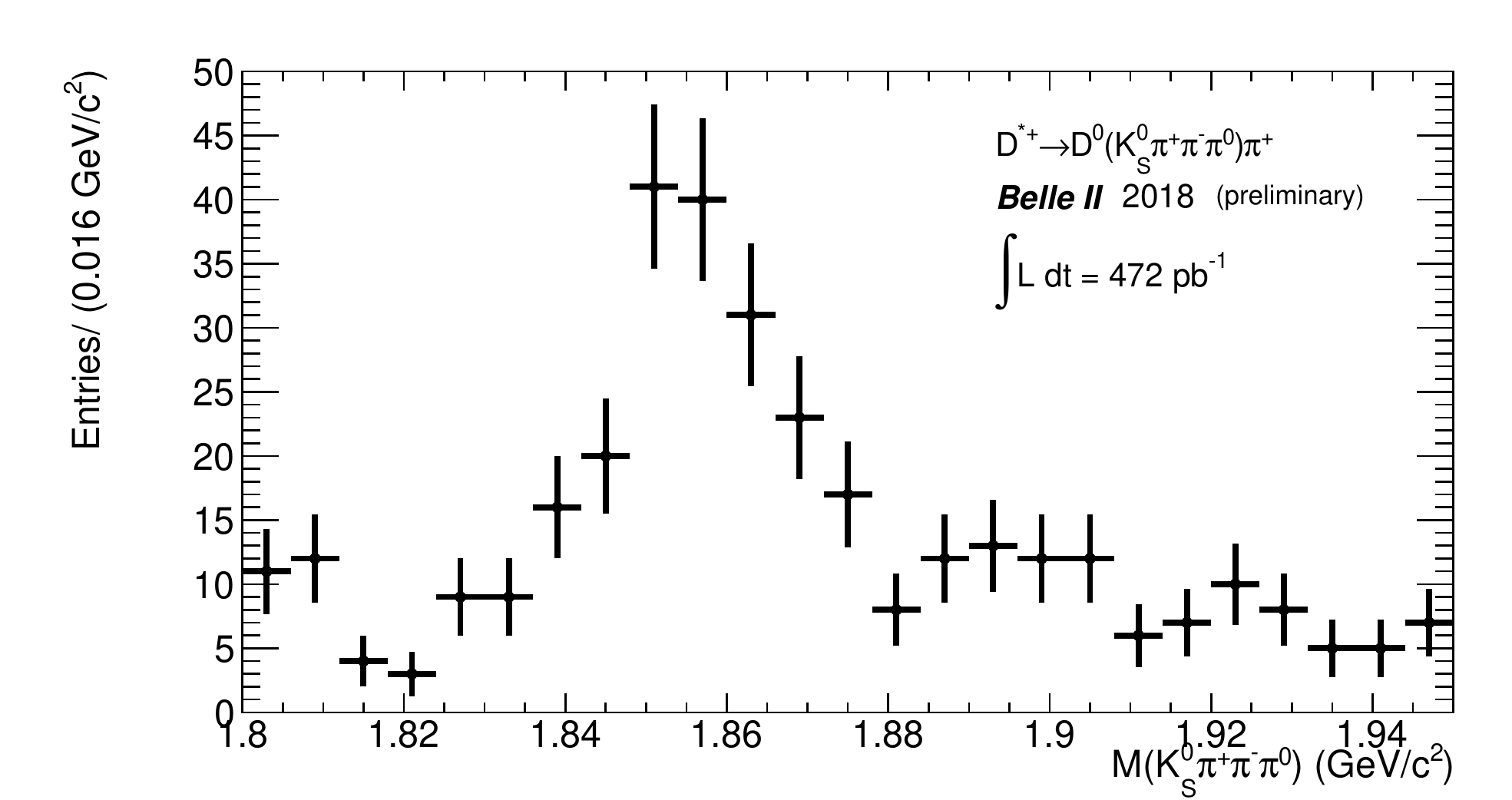} \\
\end{tabular}
\caption{$\Delta M$ (left) and $M_{D}$ (right) distributions for $D^{*\pm} \to D(K_{\rm S}^{0}\pi^+\pi^-\pi^{0})\pi^{\pm}_{\rm{slow}}$ decays.}\label{Fig:kspipipi0}
\end{figure}

 The multibody final states $K_{\rm S}^0 \pi^+\pi^-$ and $K_{\rm S}^0 \pi^+\pi^- \pi^0$ have also been rediscovered. Their $\Delta M$ and $M_{D}$ distributions are given in Fig.~\ref{Fig:kspipi} and \ref{Fig:kspipipi0}.

This establishes the capabilities of Belle~II detector to reconstruct various final state particles including the neutral ones.

\begin{figure}[ht!]
\begin{tabular}{cc}
\includegraphics[width=0.5\columnwidth]{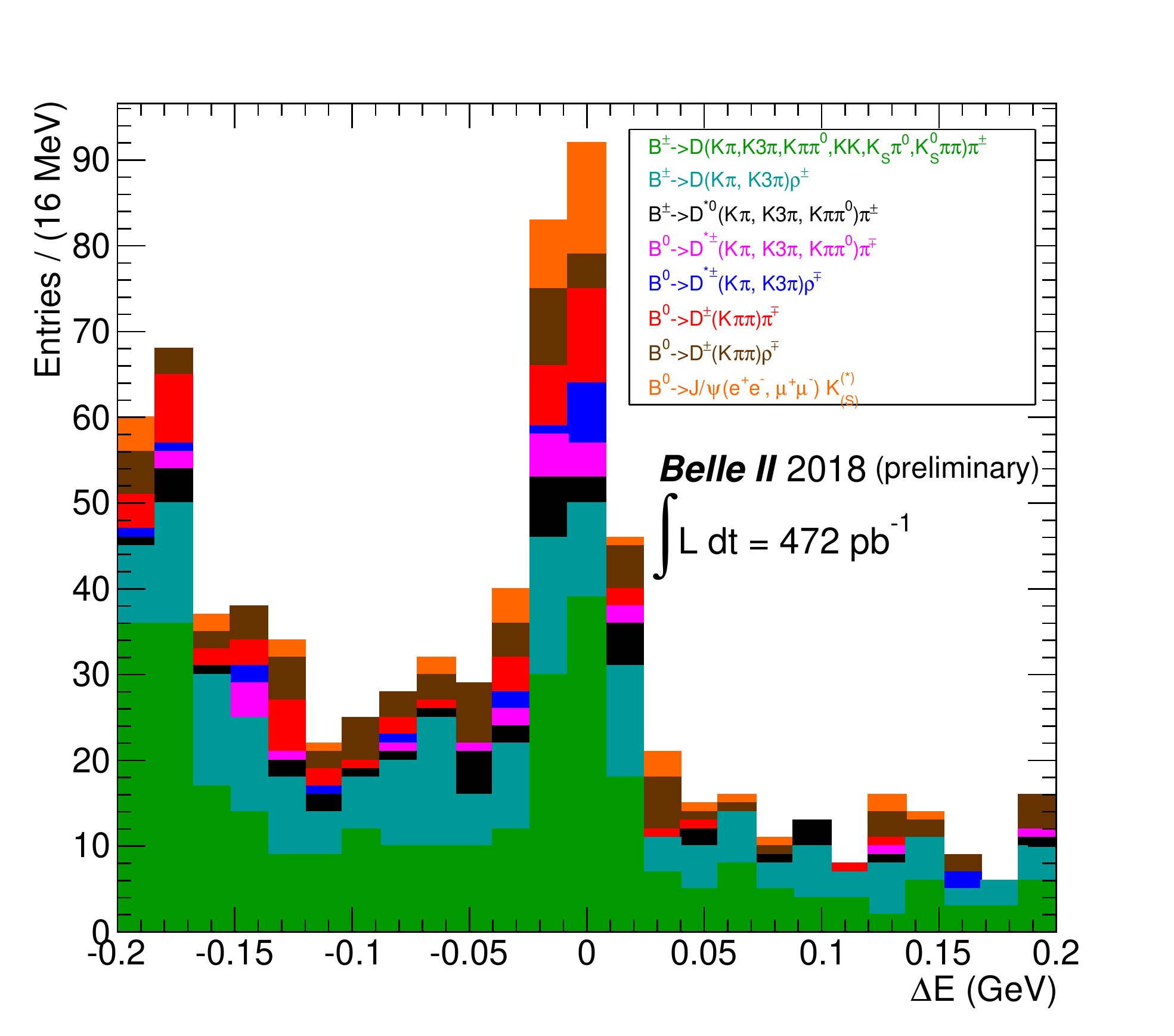} &
\includegraphics[width=0.5\columnwidth]{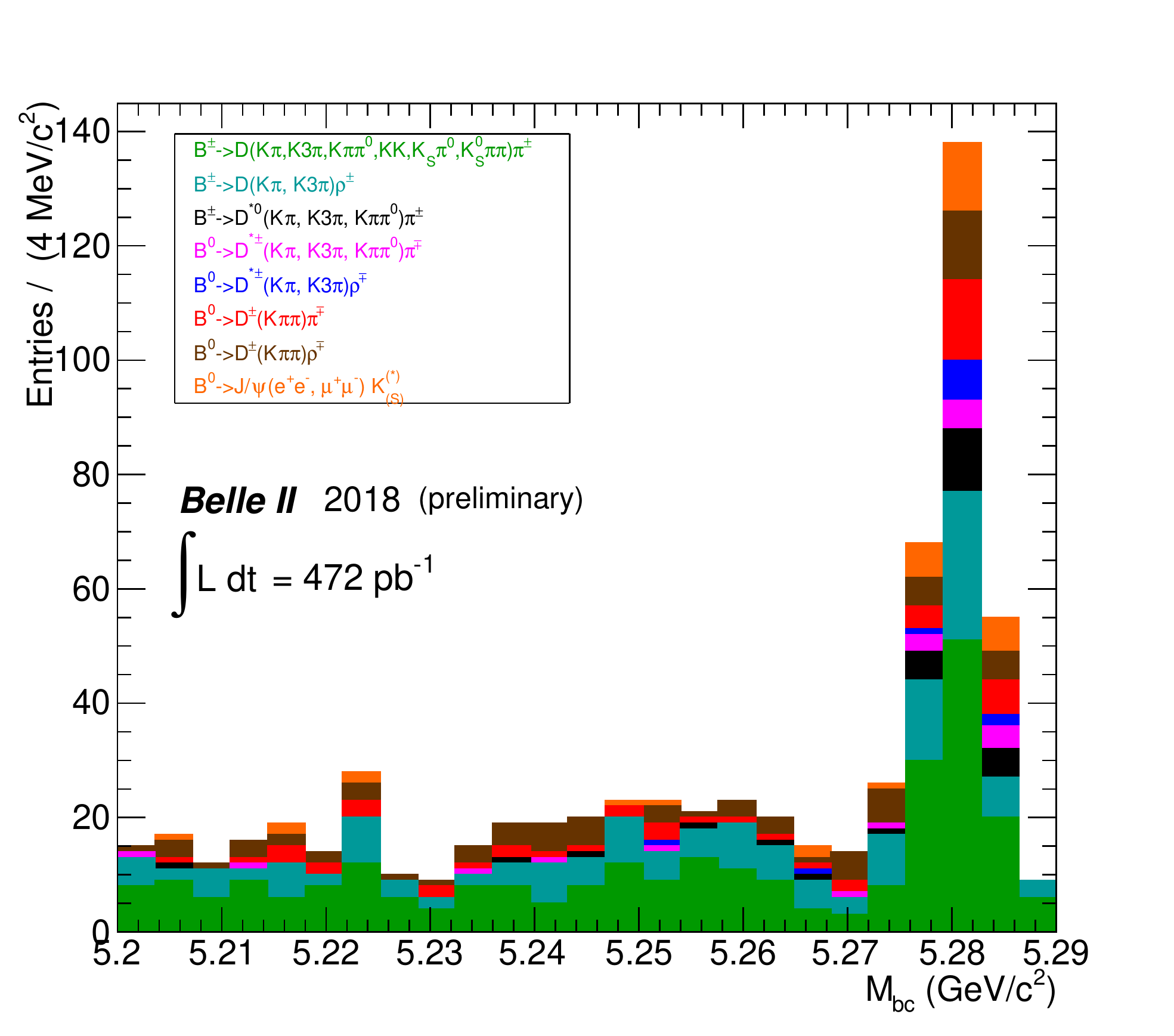} \\
\end{tabular}
\caption{$\Delta E$ (left) and $M_{bc}$ (right) distributions for various $B$ decay modes with early Belle II data.}\label{Fig:B}
\end{figure}

Around 245 $B$ candidates have been observed in various decay modes, mostly in $B\to D\pi$ which are good calibration modes for $B\to DK$. The $\Delta E$ and $M_{bc}$ distributions are shown in Fig.~\ref{Fig:B}, where $M_{\rm{bc}} = \sqrt{(E_{\rm{beam}}^{*}/c^{2})^{2} - (p_{B}^{*}/c)^{2}}$ with $p_{B}^{*}$ being the momentum of the  $B$ meson candidate in centre of mass frame.

\section{Summary}

The precise measurement of $\phi_3$ is important to establish the standard model description of $CP$ violation. It is important to add more $D$ final states for this purpose. Belle has established the standard GLW, ADS and GGSZ measurements with its full dataset. The currently ongoing analyses aim at adding unexplored and challenging decay channels to improve the overall precision. The upcoming Belle II experiment shows strong prospects for $\phi_3$ measurements. The combined sensitivity is expected to reach 1$^{\circ}$ with the full 50~ab$^{-1}$ data at Belle II. First $e^+e^-$ collisions were recorded at Belle II without the vertex detector during April-July 2018. Physics run with the full Belle II detector is expected to kick-off in early 2019.

\end{document}